\documentclass[twocolumn,preprintnumbers,amsmath,amssymb]{revtex4}

\usepackage{graphicx}
\usepackage{dcolumn}
\usepackage{bm}

\begin{document}

\title{\bf First-principles extrapolation method for accurate CO
adsorption energies on metal surfaces }

\author{Sara E. Mason, Ilya Grinberg and Andrew M. Rappe}
\affiliation{ The Makineni Theoretical Laboratories, Department of
Chemistry\\ University of Pennsylvania, Philadelphia, PA 19104-6323 }%

\date{\today}

\begin{abstract}
We show that a simple first-principles correction based on the difference between the singlet-triplet CO excitation energy values obtained by DFT and high-level quantum chemistry methods yields accurate CO adsorption properties on a variety of metal surfaces.
  We demonstrate a linear relationship between the CO adsorption energy and the CO singlet-triplet splitting, similar to the linear dependence of CO adsorption energy on the energy of the CO 2$\pi$* orbital found recently {[Kresse {\em et al.}, Physical Review B {\bf 68}, 073401 (2003)]}.  Converged DFT calculations underestimate the CO singlet-triplet excitation energy $\Delta E_{\rm S-T}$, whereas coupled-cluster and CI calculations reproduce the experimental  $\Delta E_{\rm S-T}$.
The dependence of $E_{\rm chem}$ on $\Delta E_{\rm S-T}$ is used to extrapolate $E_{\rm chem}$
 for the top, bridge and hollow sites for the (100) and (111) surfaces of Pt, Rh, Pd and Cu to the values that correspond to the coupled-cluster and CI $\Delta E_{\rm S-T}$ value.  The correction reproduces experimental  adsorption site preference for all cases and obtains $E_{\rm chem}$ in excellent agreement with experimental results.

\end{abstract}

\maketitle

\section{\label{sec:level1}Introduction }

The chemisorption of carbon monoxide  on transition metal surfaces is regarded
as a prototypical system for the study of molecule-surface interactions and has been intensely studied theoretically and experimentally~\cite{Ge00p1}.  For first-principles theoretical studies, density functional theory (DFT) with the generalized gradient approximation (GGA) to the exchange-correlation functional has emerged as the method of choice, and DFT-GGA studies have significantly advanced understanding of surface phenomena~\cite{Gross,Greeley02p319}.

Despite these successes, two of the most basic properties of CO-metal surface interactions -- the chemisorption energy and the preferred adsorption site, cannot be reliably predicted by DFT calculations.  Theoretical CO chemisorption energies obtained by the widely used PW91~\cite{Perdew92p6671} and PBE~\cite{Perdew96p1227} functionals are significantly higher than experimental values, sometimes by as much as 0.4 eV~\cite{Hammer99p7413} (30$\%$).  The RPBE functional~\cite{Hammer99p7413} does improve the adsorption energetics, but at the expense of lower accuracy in metal lattice constants and a severe underestimation of surface energies.  Even more importantly, neither PBE nor RPBE calculations can correctly predict the relative energetics of the high symmetry sites, favoring the more coordinated bridge and hollow sites over the top site, and resulting in a wrong site preference in a number of systems.  This was first discussed  by Feibelman {\em et al.}~\cite{Feibelman01p4018} where a variety of DFT-GGA methodologies and codes predicted adsorption at 
the fcc or hcp hollow site to be preferred over the experimentally preferred top site adsorption on the 
Pt(111) surface at low coverage.  Since then this ``puzzle'' has been 
addressed~\cite{Grinberg02p2264,Olsen03p119,Gil03p71,Kresse03p0734014} and the DFT-GGA inaccuracy traced to the incorrect description of CO electronic structure and bond-breaking. 

Grinberg {\em et al.} showed that the inaccuracy in site preference was due to the poor treatment of CO bond breaking by GGA functionals.
Analyzing the CO electronic structure, Gil {\em et al.} found that PBE and B3LYP~\cite{Becke93p5648,Lee88p3098} functionals place the unfilled CO 2$\pi$* orbital too low in energy.  This makes it too close in energy to the metal $d$-band, which results in an unrealistic strengthening of the 2$\pi$*- $d$-band bonding interaction.  More recently, Kresse and co-workers~\cite{Kresse03p0734014} used DFT calculations of varying accuracy and 
showed that a linear relationship exists between the difference of top and 
hollow site chemisorption energies for CO on Pt(111) and the gas-phase energy of the CO 2$\pi$* orbital.







  By using a GGA + U type functional, Kresse {\rm et al.}~\cite{Kresse03p0734014} were
able to adjust the energetic position of the gas-phase CO 2$\pi$* orbital, restoring
the correct prediction that top site adsorption is preferred on Pt(111).
However, the ``correct''  value of U is not known {\em a priori}.  Furthermore, they studied adsorption on Pt(111) and did not address CO on other metal surfaces.
  In this paper, we demonstrate  that a linear relationship  exists between the CO chemisorption energy  and the CO  singlet-triplet excitation energy for the top, bridge and hollow sites on a variety of metal surfaces.   Unlike the energetic position of the 2$\pi$* orbital, the CO  singlet-triplet excitation energy is rigorously well-defined and is accurately computed by coupled-cluster~\cite{Zeiman03p131} and configuration interaction (CI) quantum chemical calculations~\cite{Talbi00p5872}.  Extrapolation of $E_{\rm chem}$ values to the correct  CO  singlet-triplet excitation energy relies only on first-principles calculations, and yields chemisorption and site-preference energies in excellent agreement with experiment for all systems studied.

\section{\label{sec:level1} Methodology }

Calculations are carried out using the PBE GGA exchange-correlation
functional~\cite{Perdew96p1227} and norm-conserving optimized
pseudopotentials~\cite{Rappe90p1227} with the designed nonlocal
method~\cite{Ramer99p59} for metals.
All pseudopotentials were constructed using the OPIUM pseudopotential
package~\cite{Opium}.  CO chemisorption is modeled at 1/4 monolayer coverage on five layer slabs, separated by vacuum, with allowed relaxation in the top two layers.
All calculations are done, and tested to be converged,
using a $4 \times 4 \times 1$ grid of Monkhorst-Pack
$k$-points~\cite{Monkhorst76p5188}.  The Kohn-Sham orbitals are
expanded in a plane-wave basis set truncated at either 81 or 50 Ry,
the higher cutoff being required in calculations using C and O
pseudopotentials with small real-space cutoffs.  We calculate the
chemisorption energy for CO adsorbed on the (111) and (100) surfaces
of Pt, Rh, Pd, and Cu, in three high symmetry sites: top, bridge, and
hollow.  On the (111) surfaces, adsorption at the fcc hollow site is
not reported, since it is well known that differences between
calculated fcc and hcp hollow site adsorption are
negligible~\cite{Koper00p4392,Loffreda99p68}.

We repeat these calculations using different sets of C and O
pseudopotentials described in Table I.
We use three sets of C and O
pseudopotentials to calculate $E_{\rm chem}$ at each site on each
surface.  To evaluate whether three data points are adequate to
describe trends in the chemisorption energies, we expand the number of
C and O pseudopotential sets to five, and repeat the calculations for
the hollow site on Pt(111).  The slope for $E_{\rm chem}^{\rm GGA}$
versus $\Delta E_{\rm S-T}$ is unchanged on going from three points to
five, as is the goodness of fit.  Based on this, the remainder of our
results use the first three pseudopotential sets.

\section{\label{sec:level1} Results }

  For CO on Pt(111), our calculations show a linear relationship between $E_{\rm chem}^{\rm hcp}$ and CO $\Delta E_{\rm S-T}$ excitation energy (Figure 1), similar to the linear relationship between  CO/Pt(111) $\Delta E_{\rm top-fcc}$ site preference energy and the energy of the CO 2$\pi$*   for the Pt(111) adsorption observed by Kresse and coworkers~\cite{Kresse03p0734014}. A linear fit also describes well the behavior of chemisorption energies on the seven other substrates included in the present work.
Since the CO triplet state is produced by an excitation of an electron from the 5$\sigma$ orbital to the 2$\pi$* orbital, the singlet-triplet excitation energy is closely related to the 5$\sigma$-2$\pi$* gap and to the position of the 2$\pi$* orbital.    This gives rise to a similar linear dependence of $E_{\rm chem}$ on the energy of CO 2$\pi$* orbital and on $\Delta E_{\rm S-T}$. 

Coupled-cluster~\cite{Zeiman03p131} and CI~\cite{Talbi00p5872} quantum chemical calculations accurately reproduce the experimental $\Delta E_{\rm S-T}$ of 6.095~eV~\cite{Herzberg}.  On the other hand, regardless of the pseudopotential set, our DFT-GGA calculations always predict a $\Delta E_{\rm S-T}$ that is too small.  
The correct chemisorption energy $E_{\rm chem}^{\rm corr}$ can be obtained by using the relationship between $E_{\rm chem}$ and $\Delta E_{\rm S-T}$ and extrapolating to the CI  $\Delta E_{\rm S-T}$ value: 
\begin{eqnarray}
 E_{\rm chem}^{\rm corr} = E_{\rm chem}^{\rm GGA} + (\Delta E_{\rm S-T}^{\rm CI} - \Delta E_{\rm S-T}^{\rm GGA}) \frac {\delta E_{\rm chem}^{\rm GGA}} {\delta \Delta E_{\rm S-T}^{\rm GGA} },
\end{eqnarray}

\noindent where  $\Delta E_{\rm S-T}^{\rm CI}$ and $\Delta E_{\rm S-T}^{\rm GGA}$ are respectively the CI and GGA CO singlet-triplet excitation energies,  and $\delta E_{\rm chem}^{\rm GGA}$ / $\delta \Delta E_{\rm S-T}^{\rm GGA}$  is the slope of the fit  of $E_{\rm chem}$ versus $\Delta E_{\rm S-T}$.

A universal feature of the corrected chemisorption energies in Table II is 
that they all indicate weaker chemisorption than the corresponding
uncorrected values, with the $E_{\rm chem}^{\rm corr}$  values for the preferred site demonstrating much better agreement with experimentally
determined adsorption energies.  For example, the 
$E_{\rm chem}^{\rm GGA}$ for the experimentally seen CO/Pd(111) hollow site is 1.96
eV, as compared to 1.47-1.53 eV obtained by temperature programmed desorption (TPD) measurements~\cite{Conrad74p462}. This
rather large 0.46 eV error is eliminated by the use of the
extrapolation, with $E_{\rm chem}^{\rm corr}$ of 1.60~eV, in
very close agreement with experimental results.  For the experimentally
observed top site on Cu(111) the results are less dramatic but still
noticeable, with $E_{\rm chem}^{\rm GGA}$ of 0.746~eV
changed to $E_{\rm chem}^{\rm corr}$ of 0.621~eV in better agreement with the experimental value of 0.49~eV~\cite{Kirstein86p505}.  Overall, comparison of PBE and corrected results shows an improvement from 0.38~eV (30$\%$) average overestimation for the PBE functional to 0.16~eV (13$\%$) average error for our corrected results.  

Examination of the data in Table II shows that there is a strong correlation
between the magnitude of the correction and the chemisorption site.
  This is due to the different strengths of the metal-CO
interactions in different local geometries.  While the chemisorption bond is formed through both
$\sigma$ donation and $\pi$* back-donation, the contribution of $\pi$*
back-donation to the adsorption bonding in the systems considered here
is dominant because the fillings of the late
transition metals studied here are greater than half~\cite{Hammer96p2141}.  The
back-donation mechanism is strongly enhanced by going from top site to
polycoordinated adsorption sites~\cite{Illas95p12372,Koper00p4392,Kresse03p0734014,Blyholder64p2772}.  Accordingly, the  incorrect DFT-GGA $\Delta E_{\rm S-T}$ (or the incorrect placement of the 2$\pi$* orbital)   will require the smallest
correction for $E_{\rm chem}$ of the top site, followed by the bridge site, the
three-fold hcp hollow site on the (111) surfaces and the four-fold hollow
on the (100) surfaces.  This ranking is evident in the tabulated
results.  In the chemical language, the CO bond weakening is smallest
for the top site and largest for the four-fold hollow site.  

The DFT-GGA errors in the prediction of the preferred site are a
direct outcome of the unequal treatment of the CO bond weakening at
the top and hollow sites due to the unrealistically small $\Delta
E_{\rm S-T}$ and low 2$\pi$* energy.  Using our first-principles
extrapolation to eliminate the CO bond weakening errors, our corrected
DFT results give the highest value of $E_{\rm chem}^{\rm corr}$ for
the experimentally observed sites in all cases.  For Pt(111), our raw
DFT data show an incorrect site preference with an energy of 0.076~eV.  The corrected energies agree with experimental site preference~\cite{Ertl77p393,Steininger82p264,Yeo97p392} with an energy difference of 0.163~eV.  Likewise, our raw DFT data for Rh(111) is disparate with experimental site preference~\cite{Peterlinz91p6972,Hopstaken00p5457} while our corrected energies are in agreement.  For Cu(111) the use of the correction gives the experimental site preference~\cite{Hollins79p486,Kirstein86p505}, though the corrected DFT results predict small ($\leq$ 0.1~eV) differences between the top, bridge and hollow sites.  Both the raw DFT and the corrected results for Pd(111) agree with the site preference observed by experiment~\cite{Rose02p48,Guo89p6761,Conrad74p462}.  
Our corrected results for the (100) surfaces agree with experimentally observed preferred adsorption sites for Pt~\cite{Curulla00p101,Thiel83p7448,Yeo97p1990}, Rh~\cite{Kose99p8722,Medvedev97p341}, Pd~\cite{Yeo97p1990,Tracy69p4852}, and Cu~\cite{Tracy72p2748,Truong92pL385}.


A less exact but simpler correction can be extracted from our
data and applied to any CO/metal surface system.  The data in Table II
show that for any given site the $\delta E_{\rm chem}^{\rm GGA}$ / $\delta \Delta E_{\rm S-T}^{\rm GGA}$ values 
 and the consequent correction energy
are fairly constant with scatter of about 0.1~eV across a range of systems.  

As discussed above, the similarities for the
same site and the differences among the sites are consequences of
the different degrees of CO bond weakening.  The
degree of CO bond weakening can be estimated from the frequency shift
of adsorbed CO relative to the gas phase CO molecule, which  can be
easily calculated for any system.  Then the frequency can be compared
to those of top (typically 2000-2100~cm$^{-1}$), bridge (1850-1950~cm$^{-1}$),
hcp (1750-1800~cm$^{-1}$) or four-fold hollow (1600-1700~cm$^{-1}$) sites and a
corresponding correction applied. 

For example, for  CO adsorption on Ni(111), DFT-GGA calculations with PBE or PW91 functionals find the preferred site to be hcp or fcc~\cite{Eichler03p332}, in agreement with experimental results~\cite{Davis96p1367}.  However, the $E_{\rm chem}$ are in 1.9-2.0~eV range, in contrast to experimental $E_{\rm chem}$ of 1.12-1.55~eV.  Since the hollow site CO/Ni(111) stretch frequency of 1800~cm$^{-1}$~\cite{Eichler03p332} is similar to the 1830~cm$^{-1}$ frequency for  CO on Pd(111) hollow site~\cite{Loffreda99p68}, we expect the $E_{\rm chem}$ error to be similar, and the corrected value of $E_{\rm chem}$  is about 1.55~eV, in much better agreement with experimental results.  A graphical representation of the suggested correction to $E_{\rm chem}$ as a function of CO stretch frequency is given in Figure 2.

\section{\label{sec:level1} Conclusion}    

We have shown that the chemisorption energies of CO adsorbed on metal
surfaces depend linearly on gas-phase CO singlet-triplet splitting.
The difference between the high level quantum chemistry
coupled-cluster/CI and DFT-GGA singlet-triplet excitation energies can
then be used to extrapolate to chemisorption energies with the CO
error removed.  The corrected values are in good agreement with
experimental results.  The correction also eliminates the GGA errors
in site preference.  
We find a strong correlation between the amount of CO bond
breaking and the correction magnitude. This suggests that an estimate
of the GGA error due to the incorrect description of CO electronic
structure can be readily obtained through the frequency shift of
adsorbed CO at any site on a metal surface.
The demonstrated method should be applicable to
various adsorption systems where the charge transfer responsible for
chemisorption is sensitive to the adsorbate electronic
structure.  

\section{\label{sec:level1} Acknowledgments}    
This work was supported by the Air Force 
Office of Scientific Research, Air Force Materiel Command, USAF, under
grant number F49620--00--1--0170, and the NSF MRSEC Program, under Grant DMR00-79909.  
AMR acknowledges the support of the
Camille and Henry Dreyfus Foundation.  Computational support was
provided by the Defense University Research Instrumentation Program, the National Center for Supercomputing Applications and
the High-Performance Computing Modernization Office of the Department of Defense.

\begin{table}
\caption{Pseudopotential details. Core radii are in $a_o$, plane wave cutoffs $q_c$ in~Ry.  All PSPs were created from the $s^2p^2$ reference configuration for carbon and the $s^2p^4$ reference configuration for oxygen.  For each pseudopotential set, results from gas-phase molecule calculations for the 2$\pi$* energy, as well as the 5$\sigma$-2$\pi$* gap and the singlet-triplet energy are given, all in eV.}

\begin{tabular}{cccccc}

&\multicolumn{1}{c}{$r_c^O$,$r_c^C$}
&\multicolumn{1}{c}{$q_c^O$,$q_c^C$}
&\multicolumn{1}{c}{E$_{2\pi^*}$}
&\multicolumn{1}{c}{$5\sigma-2\pi^*$}&\multicolumn{1}{c}{$\Delta E_{\rm S-T}$}\\

\hline

PSP 1   &   0.94,1.09  &81,81&  -2.10  &  6.91  &  5.35\\
PSP 2   &   1.60,1.49  &47,50&  -1.90  &  7.01  &  5.53\\
PSP 3   &   1.70,1.49  &30,50&  -1.61  &  7.35  &  5.84\\
PSP 4   &   1.65,1.49  &42,50&  -1.94  &  7.04  &  5.56\\
PSP 5   &   1.70,1.49  &39,50&  -1.87  &  7.09  &  5.61\\

\end{tabular}
\end{table}

\begin{table}
\caption{Results of linear regression of chemisorption energy versus
singlet-triplet splitting energy.  The correlation coefficient for each fit is in
parenthesis next to each slope.  The DFT/GGA values for the
chemisorption energies are given, along with the corrected energies
obtained by extrapolation. For convenience, the value of the difference
between top and hollow site adsorption energies, $E_{\rm t-h}$, and 
how it evolves with
the magnitude of correction, $\Delta$, are also listed.  Positive value of $E_{\rm t-h}$ indicates that top site is preferred.  For each substrate, the site
found to be preferred by experiment is marked with an asterisk (*).
Experimental values for E$_{\rm chem}$ are given in parentheses
next to E$_{\rm GGA}^{\rm corr}$ for Pt(111)~\cite{Ertl77p393,Steininger82p264,Yeo97p392}, Rh(111)~\cite{Peterlinz91p6972,Hopstaken00p5457}, Pd(111)~\cite{Guo89p6761,Conrad74p462}, Cu(111)~\cite{Hollins79p486,Kirstein86p505}, Pt(100)~\cite{Curulla00p101,Thiel83p7448,Yeo97p1990}, Rh(100)~\cite{Kose99p8722,Medvedev97p341}, Pd(100)~\cite{Yeo97p1990,Tracy69p4852}, and Cu(100)~\cite{Tracy72p2748,Truong92pL385}  }

\begin{tabular}{rccccc}

&\multicolumn{1}{c}{Site}
&\multicolumn{1}{c}{Slope}
&\multicolumn{1}{c}{$E_{\rm chem}^{\rm GGA}$}
&\multicolumn{1}{c}{$E_{\rm chem}^{\rm corr}$}
&\multicolumn{1}{c}{$\Delta$}\\

\hline

& & & & & \\ 
Pt(111)& top* & -0.211(0.989)& 1.717& 1.560(1.43-1.71)& -0.157\\ 
& bridge& -0.435(0.999) & 1.758 & 1.433 & -0.325 \\ 
& hcp& -0.532(0.999)& 1.793 & 1.397& -0.396 \\ 
& $E_{\rm t-h}$& 0.321& -0.076& 0.164& 0.240 \\ %

& & & & & \\

Rh(111)& top* & -0.259(0.975)& 1.866 & 1.673(1.43-1.65) & -0.193\\ 
& bridge& -0.456(0.997)& 1.920 & 1.581 & -0.339\\ 
& hcp& -0.559(0.996) & 2.059 & 1.644& -0.415\\ 
& $E_{\rm t-h}$& 0.300 & -0.193 &  0.030 & 0.223\\ 
& & && & \\

Pd(111)& top& -0.185(0.977) & 1.385 & 1.247 & -0.138\\ 
& bridge& -0.399(0.977) & 1.784 & 1.487 & -0.297\\ 
& hcp*& -0.535 (0.995) & 1.962 & 1.602(1.47-1.54) & -0.360\\ 
& $E_{\rm t-h}$& 0.350& -0.577& -0.355 & 0.222\\
& & & & & \\%

Cu(111)& top* & -0.169(0.981)& 0.746 & 0.621(0.46-0.52) & -0.125\\ 
& bridge& -0.329(0.951) & 0.822 & 0.576 & -0.246\\ 
& hcp& -0.375(0.993) & 0.889 & 0.610 & -0.279\\ 
& $E_{\rm t-h}$&0.206 & -0.143 & 0.011 & 0.154\\ 
& & & & & \\%

Pt(100)& top* & -0.212(0.975)& 1.954 & 1.796(1.62-2.18) & -0.158\\ 
& bridge*& -0.422(0.996) & 2.139 & 1.824(1.62-2.18)& -.315\\ 
& hollow & -0.607(0.999) & 1.698 & 1.246& -0.452\\ 
& $E_{\rm t-h}$& 0.395& 0.256 & 0.551 & 0.295\\ 
& & & & & \\

Rh(100)& top* & -0.246(0.989) & 1.905 & 1.723(1.24-1.65) & -0.182\\ 
&bridge*& -0.427(0.999) & 2.092 & 1.774& -0.318\\ 
& hollow & -0.651(0.999) & 2.087 & 1.603 &-0.484\\ 
& $E_{\rm t-h}$ & 0.405& -0.182 & 0.120 & 0.302\\ 
& & & & & \\%

Pd(100)& top & -0.196(0.980) & 1.494 & 1.348 & -0.146\\ 
& bridge*& -0.384(0.995) & 1.927 & 1.641(1.3-1.71) &-0.286\\ 
& hollow & -0.583(0.999) & 1.937 & 1.503 & -0.434\\ 
& $E_{\rm t-h}$& 0.387 & -0.443 & -0.155 & 0.288\\ 
& & & & & \\

Cu(100)& top* & -0.170(0.996)& 0.830& 0.703(0.55-0.57) &-0.147\\ 
& bridge & -0.286(0.991) & 0.834 & 0.620 & -0.214\\ 
& hollow & -0.523(0.999) & 0.831 & 0.441 & -0.477\\ 
& $E_{\rm t-h}$ & 0.353 & -0.001& 0.262& 0.263\\

\end{tabular}
\end{table}

\clearpage
\begin{figure}
\includegraphics[width=6.0in]{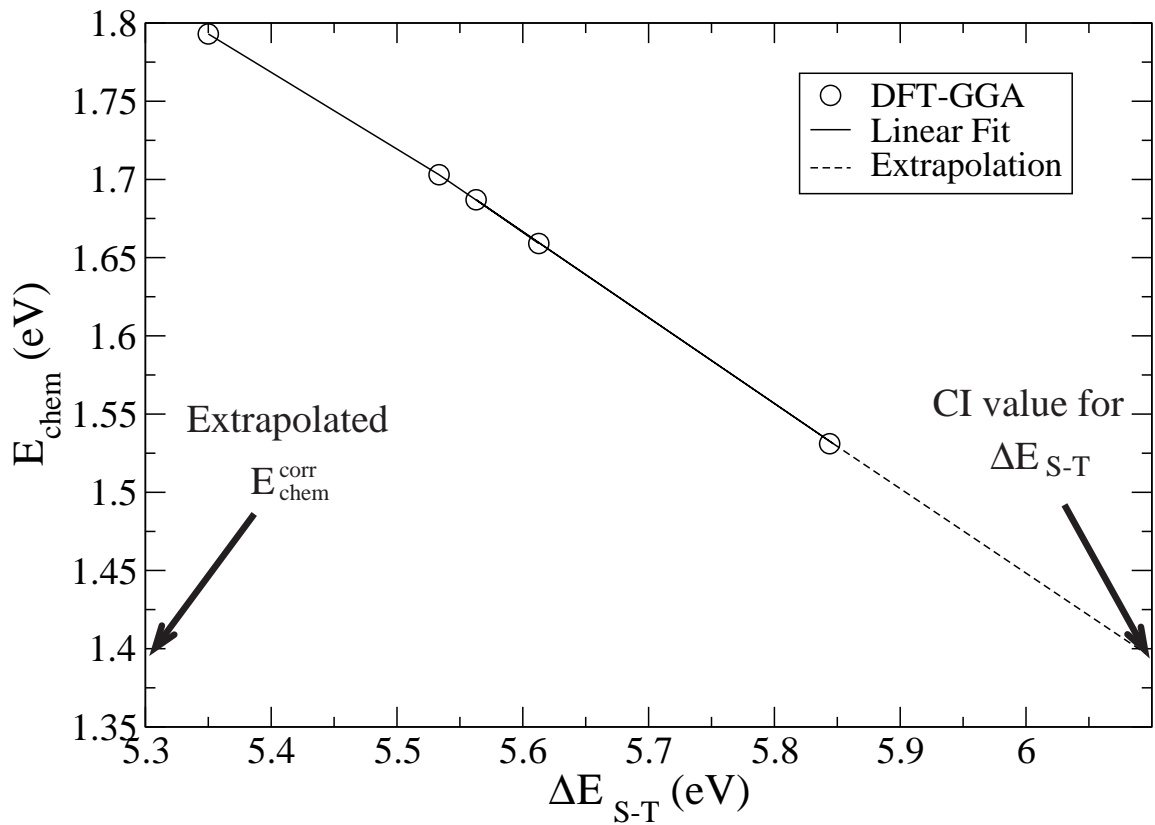}
\caption{{
$E_{\rm chem}^{\rm hcp}$ for CO on Pt(111) versus  $\Delta E_{\rm S-T}$ of CO for five pseudopotential sets (circles) and as determined by CI calculations (square).  Similar clear linear relationships are also obtained for other sites and metal surfaces.  For both the most accurate pseudopotential set as well as for the CI data point, dashed lines guide the eye to the corresponding values of $E_{\rm chem}^{\rm hcp}$.
First-principles extrapolation procedure:  $E_{\rm chem}^{\rm GGA}$ and $\Delta E_{\rm S-T}^{\rm GGA}$ values are plotted and fit to a line (solid).  This line is extended to the abscissa representing $\Delta E_{\rm S-T}^{\rm CI}$.  The corresponding ordinate value is $\Delta E_{\rm chem}^{\rm corr}$.  Chemisorption on the hcp site of Pt(111) is used in this example.
}}
\end{figure}


\clearpage
\begin{figure}
\includegraphics[width=6.0in]{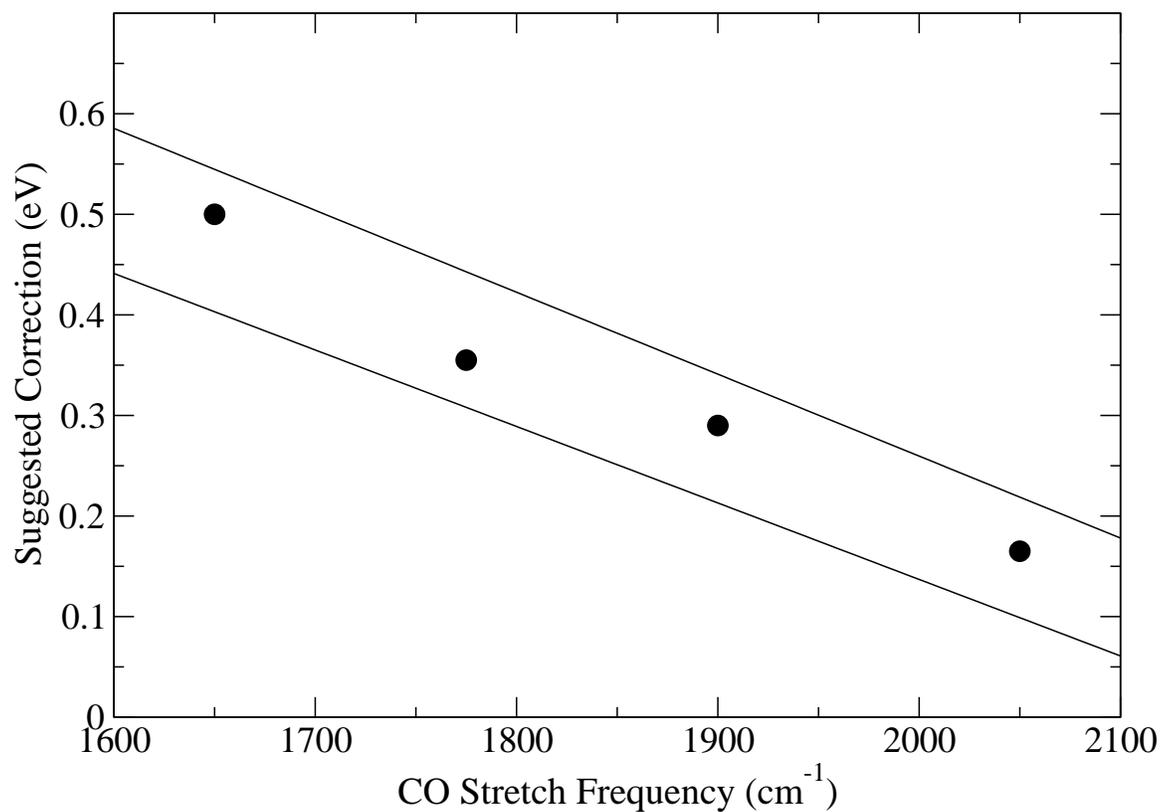}
\caption{{Graphical guide for estimating $E_{\rm chem}^{\rm GGA}$ from CO stretch frequency. Correction energy as a function of the adsorbed CO stretch frequency provides a simpler, more user-friendly  method of applying the correction.  Lines represent boundaries for suggested correction values, average correction values for high-symmetry sites for our highest quality DFT-GGA results in this study are shown as circles.}}
\end{figure} 
\clearpage
\bibliography{apssamp}

\end{document}